# Can entropy be defined for, and the Second Law applied to living systems?

**Arieh Ben Naim**

**Department of Physical Chemistry**

**The Hebrew University of Jerusalem**

**Givat Ram, Jerusalem 91904**

**Israel**

**email: ariehbennaim@gmail.com**

## Abstract

This article provides answers to the questions posed in the title. Contrary to the most common views, we show that neither the entropy, nor the Second Law may be used for either living systems, or to life phenomenon.

## 1. Introduction

As can be seen from the title, this article consists of two distinguishable questions: The first, concerns the possibility of defining entropy, and the second, concerns the applicability of the Second Law. In most cases, writers intertwine the two concepts on entropy and the Second Law[1,2]. In fact, most people *define* the Second Law as "the law of increasing entropy."[1-2] This is unfortunately, not true. One can define, interpret, and apply the concept of entropy without ever mentioning the Second Law. By the same token, one can also formulate, interpret and apply the Second Law without mentioning the concept of entropy[3,5].

In section 2, we briefly present the relatively recent definition of entropy based on the Shannon measure of information (SMI),[6-8] as well as few possible formulations of the Second Law. We show that entropy is involved in the formulation of the Second Law only for processes in isolated systems. The most general formulation of the Second Law does not depend on, nor does it use the concept of entropy.[3-4]

In section 3, we discuss the first question posed in the title of this article: Can entropy be defined for living systems? The answer to this question is a definite, No! We shall provide an argument based on SMI supporting this conclusion.

In section 4, we examine the second question posed in the title; whether or not the Second Law can be applied to living systems. We show that contrary to the opinion of a majority of writers on this subject, the Second Law of Thermodynamics *is not applicable* to any living system. This is true to any formulation of the Second Law.

In section 5, we critically examine a few quotations from the literature where entropy and the Second Law are applied to life phenomena.

## 2.   Definition of entropy based on Shannon Measure Information and various formulations of the Second Law

In this section, we very briefly present the relatively new definition of entropy based on the SMI. More details are provided in elsewhere.[3-8]

We start with the SMI, as defined by Shannon.[6] The SMI is defined for any given probability distribution $p_1, p_2, \ldots, p_N$ by:

$$H = -K \sum p_i \log p_i \qquad (2.1)$$

Here $K$ is a positive constant and the logarithm was originally taken to the base 2. Here, we use the natural logarithm, and include into $K$ the conversion factor between any two bases of the logarithm.

Clearly, the SMI defined in eq. (2.1) is a very general quantity. It is defined for any probability distribution; it can be the probabilities of Head and Tail for tossing a coin, or the six outcomes of a die. It is unfortunate that because of the formal resemblance of (2.1) to Gibbs entropy, the SMI is also referred to as entropy.[9,10,11] This has caused great confusion in both information theory and thermodynamics.[12] This confusion was recognized by Jaynes who initially adopted the term "entropy" for the SMI, but later realized the potential confusion it could create.[9,10,13]

In order to obtain the thermodynamic entropy *S* from the SMI, we have to start with the SMI and proceed in two steps: First, apply the SMI to the probability distribution of *locations* and *momenta* of a system of many particles. Second, calculate the maximum of the SMI over all possible such distributions. We note here that in this derivation we use the *continuous* analogue of the SMI written as:

$$\text{SMI(locations and velocities)} = -K \int f(\boldsymbol{R}, \boldsymbol{v}, t) \log f(\boldsymbol{R}, \boldsymbol{v}, t) d\boldsymbol{R} d\boldsymbol{v} \quad (2.2)$$

However, in actual applications for thermodynamics we always use the discrete definition of the SMI as shown in (2.1).

The procedure of calculating the distribution that maximizes the SMI in (2.2) is known as the MaxEnt (maxium entropy) principle.[10-11] We shall refer to this procedure as the MaxSMI, and not as MaxEnt.[14]

For a system of non-interacting particles the distribution of locations is uniform and that of velocities (or momenta) is Maxwell Boltzmann.[3-5, 12]

After finding the distribution that maximizes SMI in (2.2), denoted by $f^*(\boldsymbol{R}, \boldsymbol{v})$, we can calculate the maximum SMI for that particular distribution, i.e.

$$\text{MaxSMI} = -K \int f^*(\boldsymbol{R}, \boldsymbol{v}) \log f^*(\boldsymbol{R}, \boldsymbol{v}) d\boldsymbol{R} d\boldsymbol{v} \quad (2.3)$$

Once we calculate the MaxSMI for an ideal gas, we find that the value of the MaxSMI is the same as the entropy of an ideal gas as calculated by Sackur and Tetrode,[15-18] which is the entropy of an ideal gas at a specified $E$, $V$ and $N$ at equilibrium. Therefore, we can *define* the entropy of an ideal gas up to a multiplicative factor $K$, as the MaxSMI as defined in (2.3). Note that unlike the distribution $f(\boldsymbol{R}, \boldsymbol{v}, t)$ in eq. (2.2), the distribution which maximizes the SMI, denoted $f^*(\boldsymbol{R}, \boldsymbol{v})$ is not a function of time.

We now describe very briefly the procedure of obtaining the entropy from the SMI. We start with SMI which is definable on any probability distribution. We apply the SMI to two specific molecular distributions; the locational and the momentum distribution of all particles. Next, we calculate the distribution which maximizes the SMI. We refer to this distribution as the *equilibrium* distribution. Finally, we apply two corrections to the SMI, one due to the Heisenberg uncertainty principle, the second due to the indistinguishability of the particles. The resulting SMI is, up to a multiplicative constant, equal to the entropy of the gas, as calculated by Sackur and Tetrode based on Boltzmann definition of entropy.[15-19]

In a previous publication,[3] we discussed several advantages to the SMI-based definition of entropy. For our purpose in this article the most important aspect of this definition is the following:

The entropy is *defined* as the *maximum* value of the SMI, therefore entropy may be applied for equilibrium states only. This excludes any living system.

## 2.1. The Locational SMI of a Particle in a 1D Box of Length L

Suppose we have a particle confined to a one-dimensional (1D) "box" of length L. Since there are infinite points in which the particle can be within the interval (0, L), the corresponding locational SMI must be infinity. However we can define, as Shannon did, the following quantity by analogy with the discrete case:[6,12]

$$H[f] = -\int f(x)\log f(x)dx \qquad (2.4)$$

Since this quantity might either converge or diverge, we shall use only differences of this quantity for all intents and purposes. It is easy to calculate the density which maximizes the locational SMI, $H(f)$ in (2.4) which is[5,8,18]

$$f_{eq}(x) = \frac{1}{L} \qquad (2.5)$$

The use of the subscript eq (for equilibrium) will be cleared later, and the corresponding SMI calculated by (2.4) is:

$$H_{max}(\text{locations in } 1D) = \log L \qquad (2.6)$$

We acknowledge that the location of the particle cannot be determined with absolute accuracy, i.e. there exists a small interval, $h_x$ within which we do not care where the particle is. Therefore, we must correct equation (2.6) by subtracting $\log h_x$. Thus, we write instead of (2.6):

$$H(\text{location in } 1D) = \log L - \log h_x \qquad (2.7)$$

We recognize that in (2.7) we effectively defined SMI for a finite number of intervals; $n = L/h$. Note, that for $h_x \to 0$, $H$ in (2.7) diverges to infinity. Here, we do not take the mathematical limit, but we stop at $h_x$ small enough but not zero. Note also that in writing (2.7) we do not have to specify the units of length, as long as we use the same units for L and $h_x$.

## 2.2. The Velocity and momentum SMI of a Particle in 1D "Box" of Length L

Next, we calculate the probability distribution that maximizes the SMI, subject to two conditions:

$$\int_{-\infty}^{\infty} f(x)dx = 1 \qquad (2.8)$$

$$\int_{-\infty}^{\infty} x^2 f(x)dx = \sigma^2 = constant \qquad (2.9)$$

The result is the Normal distribution:[5,8,18]

$$f_{eq}(x) = \frac{\exp[-x^2/\sigma^2]}{\sqrt{2\pi\sigma^2}} \qquad (2.10)$$

Applying this result to a classical particle having average kinetic energy $\frac{m\langle v_x^2 \rangle}{2}$, and identifying the standard deviation $\sigma^2$ with the temperature of the system:

$$\sigma^2 = \frac{k_B T}{m} \qquad (2.11)$$

We get the equilibrium velocity distribution of one particle in 1D system:

$$f_{eq}(v_x) = \sqrt{\frac{m}{2mk_B T}} \exp\left[\frac{-mv_x^2}{2k_B T}\right] \qquad (2.12)$$

Here, $k_B$ is the Boltzmann constant, $m$ is the mass of the particle, and $T$ the absolute temperature. The value of the continuous SMI for this probability density is:

$$H_{max}(\text{velocity in } 1D) = \frac{1}{2}\log(2\pi e k_B T/m) \qquad (2.13)$$

Similarly, we can write the momentum distribution in 1D, by transforming from $v_x \rightarrow p_x = mv_x$, to get:

$$f_{eq}(p_x) = \frac{1}{\sqrt{2\pi mk_B T}} \exp\left[\frac{-p_x^2}{2mk_B T}\right] \qquad (2.14)$$

and the corresponding maximal SMI:

$$H_{max}(\text{momentum in } 1 \, D) = \frac{1}{2}\log(2\pi e m k_B T) \qquad (2.15)$$

As we have noted in connection with the locational SMI, the quantities (2.6) and (2.15) were calculated using the definition of the *continuous* SMI. Again, recognizing the fact that there is a limit to the accuracy within which we can determine the velocity, or the momentum of the particle, we correct the expression in (2.15) by subtracting $\log h_p$ where $h_p$ is a small, but infinite interval:

$$H_{max}(momentum\ in\ 1D) = \frac{1}{2}\log(2\pi e m k_B T) - \log h_p \qquad (2.16)$$

Note again that if we choose the units of $h_p$ (of momentum as: $mass\ length/time$) the same as of $\sqrt{mk_B T}$, then the whole expression under the logarithm will be a pure number.

## 2.3. Combining the SMI for the Location and Momentum of one Particle in 1D System

In the previous two sections, we derived the expressions for the locational and the momentum SMI of one particle in 1D system. We now combine the two results. Assuming that the location and the momentum (or velocity) of the particles are independent events we write

$$H_{max}(\text{location and momentum}) = H_{max}(\text{location}) + H_{max}(\text{momentum})$$

$$= \log\left[\frac{L\sqrt{2\pi e m k_B T}}{h_x h_p}\right] \qquad (2.17)$$

Recall that $h_x$ and $h_p$ were chosen to eliminate the divergence of the SMI for a continuous random variables; location and momentum.

In (2.17), we assume that the location and the momentum of the particle are independent. However, quantum mechanics imposes restriction on the accuracy in determining both the location $x$ and the corresponding momentum $p_x$. In Equations (2.7) and (2.16), $h_x$ and $h_p$ were introduced because we did not care to determine the location and the momentum with an accuracy greater that $h_x$ and $h_p$, respectively. Now, we must acknowledge that nature imposes upon us a limit on the accuracy with which we can determine both the location and the corresponding momentum. Thus, in Equation (2.17), $h_x$ and $h_p$ cannot both be arbitrarily small, but their product must be of the order of Planck constant $h = 6.626 \times 10^{-34}\ J\ s$. Thus we set:

$$h_x h_p \approx h \qquad (2.18)$$

And instead of (2.17), we write:

$$H_{max}(\text{location and momentum}) = \log\left[\frac{L\sqrt{2\pi e m k_B T}}{h}\right] \qquad (2.19)$$

## 2.4. The SMI of a particle in a box of Volume $V$

We consider again one simple particle in a box of volume *V*. We assume that the location of the particle along the three axes *x*, *y* and *z* are independent. Therefore, we can write the SMI of the location of the particle in a cube of edges L, and volume *V* as:

$$H(\text{location in } 3D) = 3H_{max}(\text{location in } 1D) \qquad (2.20)$$

Similarly, for the momentum of the particle we assume that the momentum (or the velocity) along the three axes *x*, *y* and *z* are independent. Hence, we write:

$$H_{max}(\text{momentum in } 3D) = 3H_{max}(\text{momentum in } 1D) \qquad (2.21)$$

We combine the SMI of the locations and momenta of one particle in a box of volume *V*, taking into account the uncertainty principle. The result is

$$H_{max}(\text{location and momentum in } 3D) = 3 \log[\frac{L\sqrt{2\pi e m k_B T}}{h}] \qquad (2.22)$$

## 2.5. The SMI of locations and momenta of *N* indistinguishable particles in a box of volume *V*

The next step is to proceed from one particle in a box, to *N* independent particles in a box of volume *V*. Given the location $(x, y, z)$, and the momentum $(p_x, p_y, p_z)$ of one particle within the box, we say that we know the *microstate* of the particle. If there are *N* particles in the box, and if their microstates are independent, we can write the SMI of *N* such particles simply as *N* times the SMI of one particle, i.e.,

$$\text{SMI(of } N \text{ independent particles)} = N \times \text{SMI(one particle)} \qquad (2.23)$$

This equation would have been correct when the microstates of all the particles were *independent*. In reality, there are always correlations between the microstates of all the particles; one is due to *intermolecular interactions* between the particles, the second is due to the *indistinguishability* between the particles. We shall discuss these two sources of correlation separately.

*(i) correlation due to indistinguishability*

Recall that the microstate of a single particle includes the location and the momentum of that particle. Let us focus on the location of one particle in a box of volume $V$. We have written the locational SMI as:

$$H_{max}(\text{location}) = \log V \qquad (2.24)$$

Recall that this result was obtained for the *continuous locational* SMI. This result does not take into account the divergence of the limiting procedure. In order to explain the source of the correlation due to indistinguishability, suppose that we divide the volume $V$ into a very large number of small cells, each of the volume $V/M$. We are not interested in the exact location of each particle, but only in which cell each particle is. The total number of cells is $M$, and we assume that the total number of particles is $N \ll M$. If each cell can contain at most one particle, then there are $M$ possibilities to put the first particle in one of the cells, and there are $M-1$ possibilities to put the second particle in the remaining empty cells. Altogether, we have $M(M-1)$ possible microstates, or configurations for two particles. The probability that one particle is found in cell $i$, and the second in a different cell $j$ is:

$$\Pr(i,j) = \frac{1}{M(M-1)} \qquad (2.25)$$

The probability that a particle is found in cell $i$ is:

$$\Pr(j) = \Pr(i) = \frac{1}{M} \qquad (2.26)$$

Therefore, we see that even in this simple example, there is correlation between the events "one particle in $i$" and one particle in $j$":

$$g(i,j) = \frac{\Pr(i,j)}{\Pr(i)\Pr(j)} = \frac{M^2}{M(M-1)} = \frac{1}{1-\frac{1}{M}} \qquad (2.27)$$

Clearly, this correlation can be made as small as we wish, by taking $M \gg 1$ (or, in general, $M \gg N$). There is another correlation which we cannot eliminate, and is due to the *indistinguishability* of the particles.

Note that in counting the total number of configurations we have implicitly assumed that the two particles are labeled, say red and blue. In this case we count the two configurations in Figure 1a, as *different* configurations; "blue particle in cell $i$, and red particle in cell $j$," and "blue particle in cell $j$, and red particle in cell $i$."

Atoms and molecules are indistinguishable by nature; we cannot label them. Therefore, the two microstates (or configurations) in Figure 1b are indistinguishable. This means that the total number of configurations is not $M(M-1)$, but:

$$\text{number of configurations} = \frac{M(M-1)}{2} \rightarrow \frac{M^2}{2}, \text{ for large } M \qquad (2.28)$$

For very large $M$ we have a correlation between the events "particle in $i$" and "particle in $j$":

$$g(i,j) = \frac{\Pr(i,j)}{\Pr(i)\Pr(j)} = \frac{M^2}{M^2/2} = 2 \qquad (2.29)$$

For $N$ particles distributed in $M$ cells, we have a correlation function (for $M \gg N$):

$$g(i_1, i_2, \ldots, i_n) = \frac{M^N}{M^N/N!} = N! \qquad (2.30)$$

This means that for $N$ indistinguishable particles we must divide the number of configurations $M^N$ by $N!$. Thus, in general by removing the "labels" on the particles, the number of configurations is *reduced* by $N!$. For two particles, the two configurations shown in Figure 1a reduce to one as shown in Figure 1b.

Now that we know that there are correlations between the events "one particle in $i_1$", "one particle in $i_2$" ... "one particle in $i_n$", we can define the *mutual information* corresponding to this correlation.[12,18] We write this as:

$$I(1; 2; \ldots; N) = \ln N! \qquad (2.31)$$

The SMI for $N$ particles will be:

$$H(N \text{ particles}) = \sum_{i=1}^{N} H(\text{one particle}) - \ln N! \qquad (2.32)$$

For the definition of the mutual information among many random variables, see references 12,18.

Using the SMI for the location and momentum of $N$ independent particles in (2.32), we can write the final result for the SMI of $N$ indistinguishable (but non-interacting) particles as:

$$H(N \text{ indistinguishable}) = N \log V \left( \frac{2\pi m e k_B T}{h^2} \right)^{3/2} - \log N! \qquad (2.33)$$

Using the Stirling approximation for $\log N!$ in the form (note again that we use the natural logarithm):

$$\log N! \approx N \log N - N \qquad (2.34)$$

We have the final result for the SMI of $N$ indistinguishable particles in a box of volume $V$, and temperature $T$:

$$H(1,2,\dots N) = N \log \left[ \frac{V}{N} \left( \frac{2\pi m k_B T}{h^2} \right)^{3/2} \right] + \frac{5}{2} N \qquad (2.35)$$

By multiplying the SMI of $N$ particles in a box of volume $V$, at temperature $T$, by the factor $(k_B \log_e 2)$, one gets the *entropy*, the *thermodynamic entropy* of an ideal gas of simple particles. This equation was derived by Sackur, and by Tetrode, in 1912 by using the Boltzmann definition of entropy.[15,16]

One can convert this expression into the entropy function $S(E, V, N)$, by using the relationship between the total energy of the system, and the total kinetic energy of all the particles:

$$E = N \frac{m \langle v \rangle^2}{2} = \frac{3}{2} N k_B T \qquad (2.36)$$

The explicit entropy function of an ideal gas is:

$$S(E, V, N) = N k_B \ln \left[ \frac{V}{N} \left( \frac{E}{N} \right)^{3/2} \right] + \frac{3}{2} k_B N \left[ \frac{5}{3} + \ln \left( \frac{4\pi m}{3h^2} \right) \right] \qquad (2.37)$$

### (ii) Correlation due to intermolecular interactions

In Equation (2.37) we got the entropy of a system of non-interacting simple particles (ideal gas). In any real system of particles, there are some interactions between the particles. Without getting into any details on the function $U(r)$, it is clear that there are two regions of distances $0 \leq r \lesssim \sigma$ and $0 \leq r \lesssim \infty$, where the slope of the function $U(r)$ is negative and positive, respectively. See Figure 2. Negative slope correspond to repulsive forces between the pair of the particles when they are at a distance smaller than $\sigma$. This is the reason why $\sigma$ is sometimes referred to as the *effective diameter* of the particles. For larger distances, $r \gtrsim \sigma$ we observe attractive forces between the particles.

Intuitively, it is clear that interactions between the particles induce *correlations* between the locational probabilities of the two particles. For hard-sphere particles there is infinitely strong repulsive force between two particles when they approach to a distance of $r \lesssim \sigma$. Thus, if we know the location $R_1$ of one particle, we can be sure that a second particle, at $R_2$ is not in a sphere of diameter $\sigma$ around the point $R_1$. This *repulsive* interaction may be said to introduce *negative correlation* between the locations of the two particles.[12,18]

On the other hand, two argon atoms *attract* each other at distances $r \lesssim 4\text{Å}$. Therefore, if we know the location of one particle say, at $R_1$, the probability of observing a second particle at $R_2$ is *larger* than the probability of finding the particle at $R_2$ in the absence of a particle at $R_1$. In this case, we get *positive correlation* between the locations of the two particles.[12,18]

We can conclude that in both cases (attraction and repulsion) there are correlations between the particles. These correlations can be cast in the form of *mutual information* which reduces the SMI of a system of $N$ simple particles in an ideal gas. The mathematical details of these correlations are discussed in references 12,18.

Here, we show only the form of the mutual information at very low density. At this limit, we can assume that there are only *pair correlations*, and neglect all higher order correlations. The mutual information due to these correlations is:

$$I(\text{due to correlations in pairs})$$

$$= \frac{N(N-1)}{2} \int p(R_1, R_2) \log g(R_1, R_2) dR_1 dR_2 \qquad (2.38)$$

where $g(R_1, R_2)$ is defined by:

$$g(R_1, R_2) = \frac{p(R_1, R_2)}{p(R_1)p(R_2)} \qquad (2.39)$$

Note again that log g can be either positive or negative, but the average in (2.38) must be positive

## 2.6 Summary of the definition of entropy based on the SMI

We summarize the main steps leading from the SMI to the entropy. We started with the SMI associated with the *locations* and *momenta* of the particles. We calculated the distribution of the locations and momenta that *maximizes* the SMI. We referred to this distribution as the *equilibrium distribution.* Let us denote this distribution of the locations and momenta of all the particles by $f_{eq}(R, p)$.

Next, we use the equilibrium distribution to calculate the SMI of a system of $N$ particles in a volume $V$, and at temperature $T$. This SMI is, up to a multiplicative constant $(k_B \ln 2)$, identical to the *entropy* of an ideal gas at equilibrium. This is the reason we refer to the distribution which maximizes the SMI as the *equilibrium distribution*.

It should be noted that in the derivation of the entropy, we used the SMI twice; first, to calculate the distribution that maximize the SMI, then evaluating the maximum SMI corresponding to this distribution. The distinction between the concepts of SMI and entropy is essential. Referring to SMI (as many do) as entropy, inevitably leads to such an awkward statement; the maximal value of the entropy (meaning the SMI) is the entropy (meaning the thermodynamic entropy). The correct statement is that the SMI associated with locations and momenta is defined for any system; small or large, at equilibrium or far from equilibrium. This SMI, not the entropy, evolves into a maximum value when the system reaches equilibrium. At this state, the SMI becomes proportional to the entropy of the system.

Since the entropy is a special case of a SMI, it follows that whatever interpretation one accepts for the SMI will be automatically applied to the concept of entropy. The most important conclusion is that entropy is defined only for equilibrium states, it is not a function of time, it does not change with time, and it does not have a tendency to increase.

The SMI may be defined for a system with any number of particles including the case $N = 1$. This is true for the SMI. When we talk about the entropy of a system, we require that the system be very large. The reason is that only for such systems the entropy-formulation of the Second Law of thermodynamics is valid. We next turn briefly to discuss the Second Law.

### 2.7  Two formulations of the Second Law

We present here two formulations of the Second Law. The first is very common, the second is quite new.

### 2.7.1 The entropy formulation of the Second Law

The entropy formulation of the SL applies only to isolated systems. We shall formulate it for a one component system having $N$ particles. If there are $k$ components, then $N$ is reinterpreted as a vector comprising the numbers $(N_1, N_2, \ldots, N_k)$ where $N_i$ is the number of particles of species $i$.[20]

*For any isolated system characterized by the constant values of the quantities* $(E, V, N)$, *at equilibrium, the entropy is maximum over all possible constrained equilibrium states of the same system.*

### 2.7.2 The probability formulation of the Second Law

Here, we state the SL for a simple system, Figure 3 (for generalization, see reference 4). We start with a system of $N$ particles in one compartment, where $N$ is of the order of one Avogadro number, about $10^{23}$ particles. We remove the partition and follow the evolution of the system. At any point in time we define the distribution of particles by the pair of numbers $(n, N - n)$. Of course, we do not count the exact number of particles in each compartment $n$, but we can measure the density of particles in each compartment, $\rho_L = n_L/V$ and $\rho_R = n_R/V$, where $n_L$ and $n_R$ are the numbers of particles in the left (L) and right (R) compartments, respectively ($n_L + n_R = N$). From the measurement of $\rho_L$ and $\rho_R$ we can also calculate the pair of mole fractions $x_L = n_L/(n_L + n_R) = \rho_L/(\rho_L + \rho_R)$ and $x_R = n_R/(n_L + n_R) = \rho_R/(\rho_L + \rho_R)$, with $x_L + x_R = 1$. The pair of numbers $(x_L, x_R)$ is referred to as the configuration of the system. Note that the pair $(x_L, x_R)$ is also a probability distribution.

After the removal of the partition between the two compartments, Figure 3b, we can ask what the probability of finding the system with a particular configuration $(x_L, x_R)$ is. We denote this probability by $\Pr(x_L, x_R)$. Since both $x_L$ and $\Pr$ are probabilities, we shall refer to the latter as super probability; $\Pr(x_L, x_R)$ is the probability of finding the probability distribution $(x_L, x_R)$. We can now state the SL for this particular system as follows:

Upon the removal of the partition between the two compartments, the probability distribution, or the configuration, will evolve from the initial one $(x_L, x_R) = (1,0)$, (i.e. all particles in the left compartment) to a new equilibrium state characterized by a uniform locational distribution. This means that the densities $\rho_L$ and $\rho_R$ are equal (except for negligible deviations), or equivalently the mole fractions $x_L$ and $x_R$ are equal to ½ . We shall never observe any significant deviation from this new equilibrium state, not in our lifetime, and not in the universe's lifetime which is estimated to be about 15 billion years.

Note that before we removed the partition in Figure 3a, the probability of finding the configuration (1, 0) is one. This is an equilibrium state, and all the particles are, by definition, of the initial state, in the L compartment.

We refer to the super probability of finding the configuration $(x_L, x_R)$, denoted by $\Pr(x_L, x_R)$, as the probability of the configuration attained *after* the removal of the partition, when $x_L$ can, in principle, attain any value between zero and one. Therefore, the super probability of obtaining the configuration (1, 0) is negligibly small. On the other hand, the super probability of obtaining the configuration in the neighborhood of $\left(\frac{1}{2}, \frac{1}{2}\right)$ is, for all practical purposes nearly one.[8,12,18] This means that after the removal of the partition, and reaching an equilibrium state, the ratio of the super probabilities of the initial configuration (1, 0) and the final configuration, i.e. in the neighborhood of $\left(\frac{1}{2}, \frac{1}{2}\right)$, is almost infinity. With $N \approx 10^{23}$, this is an unimaginable large number). Thus, we can say that for $10^{23}$

$$\frac{\Pr(\text{final configuration})}{\Pr(\text{initial configuration})} \approx \text{infinity} \qquad (2.40)$$

This is the *probability formulation* of the Second Law for this particular experiment. This law states that starting with an equilibrium state where all particles are in L, and removing the constraint (the partition), the system will evolve to a new equilibrium configuration which has a probability overwhelmingly larger than the initial configuration.

Note carefully that if *N* is small, then the evolution of the configuration will not be monotonic, and the ratio of the super probabilities in the equation above is not near infinity. For some simulations the reader is referred to reference 21. For very large *N*, the evolution of the configuration is also not strictly monotonic, and the ratio of the super probabilities is not strictly, infinity. However, in practice, whenever *N* is large, we shall never observe any deviations from monotonic change of the configuration from the initial value (1, 0) to the final configuration $\left(\frac{1}{2}, \frac{1}{2}\right)$. Once the final equilibrium state is reached [i.e. that the configuration is within experimental error $\left(\frac{1}{2}, \frac{1}{2}\right)$], it will stay there *forever*, or equivalently it will be found with probability one.

The distinction between the strictly mathematical monotonic change and the practical change is important. The process is mathematically always reversible, i.e. the initial state will be visited. However, in practice the process is irreversible; we shall never see the reversal to the initial state.

Let us repeat the probability formulation of the Second Law for the more general process. Figure 4.

We start with an initial constrained equilibrium state, Figure 4a. We remove the constraint, and the system's configuration will evolve with probability (nearly) one, to a new equilibrium state, Figure 4b, and we shall never observe reversal to the initial state. "Never" here, means never in our lifetime, nor in the lifetime of the universe.

Note carefully that we formulated the SL in terms of the probability *without* mentioning entropy. This is in sharp contrast with most formulations of the Second Law.

So far we have formulated the Second Law in terms of probabilities. One can show that this formulation is equivalent to the entropy formulation of the Second Law. See references 8,12, 18.

The relationship between the two formulations (isolated system) is

$$\frac{\Pr(f)}{\Pr(in)} = \exp\left[\frac{\Delta S(in \to f)}{k_B}\right] \qquad (2.41)$$

Here, the probability ratio on the left hand side is the same as in eq. (2.40), except that here the initial ($in$) and the final ($f$) configurations are more general than in eq. (2.40). Note also that both brobalilities $\Pr(f)$ and $\Pr(in)$ pertain to the system *after* the removal of the partition, Figure 4c.

We now very briefly discuss two more generalizations of the Second Law. For a process at $T, V, N$ constants, we can write the relationship

$$\frac{\Pr(f)}{\Pr(in)} = \exp\left[-\frac{\Delta A(in \to f)}{k_B T}\right] \quad , \quad (T, V, \boldsymbol{N} \; system) \qquad (2.42)$$

Here, $\Delta A = \Delta E - T\Delta S$ is the change in the Helmholtz energy of the system for the general process as described in Figure 4, except that the total process is carried out at constant temperature $(T)$, rather than constant energy $(E)$.

Finally, for a process at constant $P, T, \boldsymbol{N}$ we have

$$\frac{\Pr(f)}{\Pr(in)} = \exp\left[-\frac{\Delta G(in \to f)}{k_B T}\right] \quad , \quad (T, P, \boldsymbol{N} \; system) \qquad (2.43)$$

Here, $\Delta G = \Delta A + P\Delta V$, and $\Delta V$ is the volume change in the process. Note that if we have a one-component system, and each compartment has the same $P, T$, then the chemical potential of each compartment will also be the same. Therefore, no process will occur in such a system. However, in a multi-component system, there will be a process for which $\Delta G < 0$.

Comparing equations (2.41), (2.42) and (2.43), we see that the thermodynamic formulations of the Second Law are *different* for different systems (entropy maximum for $(E, V, N)$ system, Helmholtz energy minimum for $(T, V, \boldsymbol{N})$ system and Gibbs energy minimum for $(T, P, \boldsymbol{N})$ system). On the other hand, the probability formulation is the same for all systems; the probability ratio on the left hand side of equations (2.41), (2.42) and (2.43) is of the order $e^N$. For $N \approx 10^{23}$, this is practically infinity as noted in equation (2.40).

Clearly, the probability formulation is far more general than any other thermodynamic formulation of the Second Law.

### 3. Can we define the entropy for any living system?

The answer to this question should be examined according to the various *definitions* of entropy:

(i)      Clausius' definition

Clausius did not define entropy, but rather the entropy change for one specific process. It was only much later, based on the third law of thermodynamics, that one could define an "absolute" value of entropy.

Clearly, one may apply the Clausius definition $dS = dQ/T$ for any system at constant temperature $T$. Assuming that a living system is at constant temperature, one can claim that the entropy *change* due to a flow of a small quantity of heat $dQ$ will be $dQ/T$.

However, this is quite remote from a *definition* of entropy for a living system. First, because it assumes that the only change that occurred in the system is the flow of heat. Clearly, this is not the case. Second, it is difficult to claim that a living system is a thermodynamic system at constant temperature. Since the system is open to its environment and since the temperature of the environment is not constant, the assumption of constant temperature is invalid. Besides, various processes in different cells of the body could produce heat, and can locally affect the temperature in different points in the body.

Finally, and most importantly, to obtain the entropy of the system one has to calculate the integral

$$S(T) = S(O) + \int_O^T \frac{c_p}{T} dT \qquad (3.1)$$

Leaving aside the unknown constant $S(O)$, there is no way to perform this integral for any living system. In fact, there is no way of defining the heat capacity as a function of T for any living system. Therefore, we can conclude that the Clausius definition is ruled out for this purpose.

(ii)    Boltzmann's definition

For a system having a fixed energy *E*, volume *V*, and composition $\boldsymbol{N} = (N_1, \dots, N_c)$ where $N_i$ is the number of particles of species *i*, the Boltzmann entropy is given by

$$S_B = k_B \log W \qquad (3.2)$$

Here, $k_B$ is the Boltzmann constant and *W* is the number of microstates of the system.

These microstates are essentially the stationary solution of the Schrödinger equation for a system characterized by the variables $(E, V, \boldsymbol{N})$. At this point one can say that a living system is not a well-defined thermodynamic system, and therefore we cannot apply any of the thermodynamic functions (including energy, entropy, Gibbs energy) to describe it. However, there is a deeper reason why we cannot apply the Boltzmann entropy to a living system; we do not know whether or not one can describe any living system, even the simplest one in terms of a wave function. Equivalently, we do not know whether or not Schrödinger's equation is applicable to a living system. Here, I do not mean "applicable" in the sense that the Hamiltonian operator of the system is extremely complicated, or that solving the Schrödinger equation would be extremely difficult. These difficulties are certainly true but they are also true of many well-defined thermodynamic systems. Here, I mean that a living system might, *in principle*, not be *described* by a wave function, hence there is no point in talking about the corresponding Schrödinger equation, hence calculating the number of microstates. Thus, the unavailability of $W$ is not a problem of difficulty in calculation, but rather a result of the impossibility of describing any living system by any physical theory presently available. This point has been discussed at great length by Penrose.[22] Further the discussion of this aspect of the Schrödinger equation, and the Schrödinger's cat may be found in Ben-Naim (2016).[23]

### (iii)    *The definition based on the SMI*

In a previous book,[3] it was argued that the definition of entropy based on the SMI is superior to all other definitions of entropy. One of the most important conclusions of the definition based on SMI is that entropy is obtained as the *maximum* of SMI over all possible distributions of locations and momenta. This maximum is identified with the *equilibrium* value of the SMI, and at equilibrium the entropy is proportional to the SMI. It follows that entropy is defined only for well-defined thermodynamic systems at equilibrium. Living systems are not well-defined thermodynamic systems at equilibrium, and therefore entropy cannot be assigned to any living system.

### 4.   Is the Second Law applicable to life phenomena?

Anyone who studied thermodynamics knows that there are many different formulations of the Second Law of thermodynamics (SL). These formulations were recently reviewed.[3,4] The most popular statement of the SL is:[2] "The law of increasing entropy." This formulation follows from the well-known statement made by Clausius:

"The entropy of the world always increases."

These statements were recently criticized by Ben-Naim.[3,4] The entropy by itself, has no value, and therefore cannot either increase or decrease. The "entropy of the universe" is not even defined, and therefore one cannot say whether it increases or decreases.

As we have pointed out in the previous section, no one has ever defined the entropy of any living system, and I doubt that it will ever be defined for such systems. Therefore, using the entropy formulation of the SL in the above form cannot be applied to a living system. This is also true for the correct entropy-formulation of the SL, see section 2.

Some authors who recognize that living systems are not isolated, and processes in living creatures are not carried out in isolated conditions, use one of the Free-energy formulation of the Second Law.[2] Unfortunately, if entropy may not be defined for any living system, it also follows that the Helmholtz or the Gibbs energy are not defined for any living system.

Perhaps the most convincing argument against the usage of the Second Law for life processes follows from the most general formulation of the Second Law, to which we refer to as the Probability-formulation. See section 2.

Clearly, for any system having many particles one can claim that configuration having higher probability will be observed more frequently. This is plain common sense.[24] For systems with $10^{23}$ particles, "more frequently" turns into "practically always." This is why we never observe the reversal of a spontaneous process in macroscopic systems.[21,24]

Turning to processes in living systems, in particular; thinking, feeling and creation of arts, it is difficult, if not impossible to argue that these processes occur because they follow the laws of probability.

Thus, as long as entropy was ill-understood, one could assign entropy values to a living system, then claim that the "Law of increase of entropy" applies to living systems, and includes even those activities which we believe to be a result of our free will.

However, once we grasp the idea that the Second Law is nothing but the law of probability applied to a system of large number of particles, it will be absurd to claim that what you and I think, feel or create arts is a result of the laws of probability.

We therefore conclude that the Second Law is inapplicable to processes occurring in living systems, certainly not to such processes as thinking, feeling, and creation of arts.

## 5. Examples of misuses of entropy and the Second Law in living systems

We present here a few examples of misusing entropy and the Second Law for living systems.

It is difficult to trace the origin of the application of entropy and the Second Law to living systems. There is no doubt however, that the most prominent and influential scientist who was responsible for much of the nonsensical writings about the involvement of entropy in life phenomena was Schrödinger.[25]

In his famous and widely praised book, "What is Life," he expressed several times the erroneous idea that the Second Law is the "natural tendency" of things to go from *order* to *disorder*," and in addition: "Life seems to be orderly and lawful behavior of matter, not based exclusively on its tendency to go over from order to disorder."

Schrödinger first asks:

*"How does the living organism avoid decay? The obvious answer is: By eating, drinking, breathing and (in the case of plants) assimilating. The technical term is metabolism.*"

Indeed, this is the obvious answer. However on page 76 of his book Schrödinger explains:

*"What then is that precious something contained in our food which keeps us from death? That is easily answered. Every process, event, happening – call it what you will; in a word, everything that is going on in Nature means an increase of the entropy of the part of the world where it is going on. Thus, a living organism continually increases its entropy – or, as you may say, produces positive entropy – and thus tends to approach the dangerous state of maximum entropy, which is death. It can only keep aloof from it, i.e. alive, by continually drawing from its environment negative entropy – which is something very positive as we shall immediately see. What an organism feeds upon is negative entropy. Or, to put it less paradoxically, the essential thing in metabolism is that organism succeeds in freeing itself from all the entropy it cannot help producing while alive."*

This paragraph contains several unfounded statements: First, it is not true that everything that goes on in Nature means an "increase of the entropy." Second, it is not true that living things

"produce positive entropy," Third, it is not clear what "the dangerous state of maximum entropy" (for a living organism) is, and finally, the meaningless idea that the only way a living system can stay alive is by drawing *negative entropy* from its environment.

Thus, Schrödinger not only adopted the misinterpretation of entropy as a measure of disorder, and not only expressed the misconception regarding the role of entropy in living systems, but also "invented" a new concept of "negative entropy" to explain how a living system "keeps aloof of death."

Entropy, by definition, is a positive quantity. There is no *negative entropy*, as there is no negative temperature, or negative mass or negative time. Of course, there are processes involving negative *changes* in entropy, but there exists no *negative entropy*!

No doubt Schrödinger's slip of the tongue had a detrimental effect on how scientists used the concept of entropy. While it had no effect on his reputation as a great theoretical physicist, he had inadvertently herded many others, scientists and non-scientists alike to fall into the pit created by his "negative entropy."

It is difficult to assess the extent of the negative impact of the "negative entropy" on science and on scientists, on writers and readers of popular science books. The "negative entropy" has been transformed by Brillouin (1962)[26] into "negentropy," and negentropy into information. If information is supposed to be "everything" (as encapsulated by Wheeler's slogan "it from bit"[27]), then one is seemingly handed a blank check and thus can say anything one wants about entropy and life and no one can prove him/her wrong. Serious scientists followed the lead of the master, and spewed statements ranging from "life is a constant struggle against the Second Law" to the assertion that "entropy *explains* life itself."

We emphasize again that entropy is not definable for a living system. Any statement about the entropy change in a living system is therefore, meaningless. This is *a fortiori* true when we use the meaningless "negative entropy" in connection with living systems.

Brillouin goes even further and claims that:[26]

"*If a living organism needs food, it is only for the negentropy it can get from it, and which is needed to make up for the losses due to mechanical work done, or simple degradation processes* *in living systems. Energy contained in food does not really matter: Since energy is conserved and never gets lost, but negentropy is the important factor."*

While one is still baffled with the concept of *negative entropy*, or the shorter version of *negentropy*, one may be greatly relieved to read the "explanation" (Hoffmann, 2012) that:[28]

*"Life uses a low-entropy source of energy (food or sunlight) and locally decreases entropy (created order by growing) at the cost of creating a lot of high-entropy "waste energy (heat and chemical waste)."*

Thus, in modern books the meaningless notion of *negative entropy* (or neg-entropy) is replaced by the more meaningful term of *low entropy*.

Is it meaningful to claim that we, living organisms, feed on low entropy food? If you are convinced that feeding on low entropy food is the thing that keeps you alive, you should then take your soup (as well as your coffee and tea), as cold as possible. This will assure you of feeding on food with the lowest entropy possible. You can enjoy as much (and as cold) ice cream as you wish. Forget about the *calories* of the ice cream and remember its *entropy* value (calories divided by degrees Kelvin).

While reading about low entropy food, one may wonder why food manufacturing companies are not required by law to label the "entropy value" of their products (per 100 grams or per serving) as they usually do with the energy values (in calories).

The truth is that when we feed on a low entropy food (I mean real food, like eggs, apples, etc.) we have no idea what happens to the entropy of the food, nor what effect low entropy food have in our bodies. Even if you believe that our bodies are struggling to lower their entropy (or increase their order), it is not clear that feeding on low-entropy food will help the body lower its entropy.

We feed on food for their energy-value, and not on their entropy-value, nor for their information-value. Without the energy of the chemical bonds in the molecules we eat, we could not move, nor digest food, nor think and feel and many other activities of the body.

Some authors identify entropy with information. For instance in Seife's book (2007)[29] "Decoding the Universe," the author identifies entropy with information. In fact, he even claims that thermodynamics (which includes the concept of entropy) is a special case of information theory (which includes the Shannon measure of information). If low entropy food is equivalent to high information food, why cannot we feed on *information* itself rather than get it from food?

Seife (2007)[29] dedicates a whole chapter in his book to "Life." Before he discusses life itself he makes the statement that "thermodynamics is, in truth a special case of information theory." Clearly, entropy is one concept within thermodynamics. The SMI is also one concept, though a central one, in information theory. From this it does not follow that thermodynamics is a special case of information theory.

Then he makes the statement that "all matter and energy is subject to the "laws of thermodynamics," *including us*." In my view, there is no justification to his claim that we are subject to the laws of thermodynamics, and certainly not to the (nonexistent) "laws of information."

To the question, "How, then, can life exist at all?" Seife answers:

*"On purely physical level, it is not too much of a puzzle? Just as a refrigerator can use its engine to reverse entropy locally by keeping its insides colder than the room it is in, the cell has biological engines that are used to reverse entropy – locally - by keeping the information in the cells intact.*

If entropy is identified with information, why not label all kinds of food with their *informational* value (say in bits per 100g or per serving)? Such labels, as well as the food, will not only be easier to understand, it might be also easier to digest…

Here is an example from Katchalsky (1965).[30] He writes:

*"Life is a constant struggle against the tendency to produce entropy by irreversible process. The synthesis of large and information-rich-macromolecules…all these are powerful anti-entropic force…living organism choose the least evil. They produce entropy at a minimal rate by maintaining a steady state."*

This is a beautiful statement, but devoid of any meaning. No one knows how to *define* the entropy of a living system, how much entropy is produced by a living organism. What are these powerful anti-entropic forces, and systems maintained at a steady state?

As we have mentioned earlier, at present we do not have a definition of life. For the purpose of this section we shall be more specific and say, that we do not know how to describe the *state* of a living organism. Therefore, the concept of entropy cannot be applied to a living system. This is true even if we could have somehow developed a thermodynamic theory which applies to systems far from equilibrium. Such a theory would probably not be applicable to a living organism. Therefore, all statements involving the role of entropy in living systems are at best meaningless.

Volkenstein (2009)[31] comments on the "antientropic" by saying:

*"At least we understand that life is not "antientropic," a word bereft of meaning. On the contrary, life exists because there is entropy, the export of which supports biological processes…"*

This statement is as meaningless as the concept of "antientropic" concept it is criticizing.

Perhaps, the most extreme and absurd ideas about entropy and life were expressed by Atkins. In the introduction of Atkins' book, (1984)[1] he writes:

*"In Chapter 8 we also see how the Second Law accounts for the emergence of the intricately ordered forms characteristic of life."*

Of course, this is an unfulfilled promise. The Second Law does not account for the emergence of "ordered forms characteristic of life."

At the end of Chapter 7, Atkins writes:[1]

*"We shall see how chaos can run apparently against Nature, and achieve that most unnatural of ends, life itself."*

Again, this is an impressive but an empty statement.

Furthermore, in a more recent book, Atkins (2007)[2] says:

*"The Second Law is one of the all-time great laws of science, for it illuminates why anything, from the cooling of hot matter to the formulation of a thought, happen at all" and later in the book he adds:*

*"The Second Law is of central importance…because it provides a foundation for understanding why **any** change occurs…the acts of literary, artistic, and musical creativity that enhance our culture."*

Finally, in both of the abovementioned books, Atkins writes on the Second Law: *"…no other scientific law has contributed more to the liberation of the human spirit than the Second Law of thermodynamics."*

All these quotations are extremely impressive, yet are totally empty statements. The Second Law *does not* provide *explanation* for "formulation of a thought," and certainly not the "acts of literacy, artistic and musical creativity." And of course no one can explain how the Second Law "contributed to the liberation of the human spirit?"

In my opinion, such claims not only do not make any sense, but they can actually discourage people from even trying to understand the Second Law of thermodynamics. Life phenomena involve extremely complicated processes. Everyone knows that life is a complex phenomenon, many aspects of which, such as thoughts, feelings and creativity are far from being well understood. Therefore, the undeliverable promise of explaining life by the Second Law will inevitably frustrate the reader to the point of concluding that entropy and the Second Law, as life, are hopelessly difficult to understand.

Some people also involve entropy and the Second Law with evolution.

Monod, in his book (1997)[32], "Chance and Necessity" he writes:

*"Evolution in the biosphere is therefore a necessarily irreversible process defining a direction in time – a direction which is the same as that enjoyed by the law of increasing entropy."*

I disagree with this statement. First, because evolution is not "irreversible" in the sense used in thermodynamics. Mutation which occurs spontaneously can occur in the reverse direction. The survival of the mutation or the reversed mutation ultimately depends on the changing environment. Second, I do not see any relationship between the process of evolution and the increase in entropy.

Those who claim that the Second Law prohibits evolution, or that evolution violates the Second Law, use the platitude argument that evolution proceeds towards more ordered, or more organized organism. Since entropy is interpreted as a measure of disorder, one concludes that evolution must run against the Second Law which requires that the entropy (or disorder) always increases. This argument is unfortunately erroneous.

First, the misconception that evolution thrives towards more order, or more organized. Although this is what we would have liked to believe to be true (*we* are more ordered, more organized than primitive organisms), evolution does not *necessarily* proceed towards more ordered, more organized, or more complex organism.

Second, the identification of entropy with disorder is totally unfounded. Indeed, there are examples in which increase in disorder in a system correlates with the increase in entropy – but there are also examples for which positive change in entropy correlates with increase in order.[3]

These arguments are sufficient for rejecting the claim that evolution either violates or defies the Second Law.

In an article entitled: "Entropy and Evolution," Styer (2008)[33] begins with a question, "Does the Second Law of thermodynamics prohibit biological evolution?" Then he continues to show, *quantitatively* that there is no conflict between evolution and the Second Law. Here is how he calculates the "entropy required for evolution."

*"Suppose that, due to evolution, each individual organism is 1000 times "more improbable" than the corresponding individual was 100 years ago. In other words, if $\Omega_i$ is the number of microstates consistent with the specification of an organism 100 years ago, and $\Omega_f$ is the number of microstates consistent with the specification of today's "improved and less probable" organism, then $\Omega_f = 10^{-3}\Omega_i$."*

From these two numbers he estimates the change in entropy per one evolving organism, then he estimates the change in entropy of the entire biosphere due to evolution.

His conclusion:

*"The entropy of the earth's biosphere is indeed decreasing by a tiny amount due to evolution, and the entropy of the cosmic microwave background is increasing by an even greater amount to compensate for that decrease."*

In my opinion, this *quantitative* argument is superfluous. It adds more confusion than clarification. In fact, it weakens the *qualitative* arguments which I have given above. No one knows how to calculate the "number of states" ($\Omega_i$ and $\Omega_f$) of any living organism. No one knows *what the states* of a living organism are, let alone *count* them. Therefore, the estimated change in entropy due to evolution in the entire article by Styer is meaningless.

The Second Law does not state that a system (animate or otherwise) tends to go from order to disorder. It does not state that the entropy of an open system cannot increase or decrease. It does not state anything about a system remote from equilibrium. Therefore, all these talk about life, evolution and the Second Law are superfluous.

Life does not violate the Second law, nor does it emerge from the Second Law. The Second Law does not apply to a living system.

At this stage of our knowledge of life we can be satisfied with applying the SMI to well specified distribution functions associated with a living system.

Unfortunately, we do not know whether or not the SMI or information theory can be applied to life itself. Certainly, it cannot be applied to *explain* aspects of life that are far from being understood, such as consciousness, thoughts, feeling, creativity, etc. Here again statements claiming that information theory can help us with the comprehension of these aspects of life abounds in the literature. These statements are no doubt very impressive, but unfortunately they are far from being true.

Finally, I will present an example of an abuse of the concept of entropy. In a book titled "Genetic Entropy and the Mystery of the Genome," Sanford (2005)[34] writes:

*"For decades biologists have argued on a philosophical level that the very special qualities of natural selection can easily reverse the biological effects on the Second Law of thermodynamics. In this way, it has been argued; the degenerative effects of entropy in living systems can be negated – making life itself potentially immortal. However, all of the analyses of this book contradict that philosophical assumption. Mutational **entropy** appears to be so strong within large genomes that selection cannot reverse it. This makes eventual extinction of such genomes inevitable. I have termed this fundamental problem **Genetic Entropy**. Genetic Entropy is not a starting axiomatic position – rather, it is a logical conclusion derived from careful*

*analysis of how selection really operates."*

Nothing in this entire paragraph makes any sense.

It should be stressed that my objection to the usage of entropy and the Second Law applies to the *entire* living system and the whole life phenomena. There is no objection to studying specific chemical, mechanical, or electrical processes occurring within a living system. However, phenomena involving mental or conscious activities cannot be included in such process.

## 6. Conclusion

The most unjustifiable, unwarranted and unacceptable application of the concept of entropy and the Second Law is undoubtedly in the phenomenon we call life. Here, I mean life itself, not a particular process occurring in a living system.

The application of entropy for life follows from several erroneous ideas about entropy and life

First, one adopts the interpretation of entropy as a measure of disorder. Second, one views life itself as well as evolution of life as a process towards more order, more structure, more organized, etc.

Combining these two erroneous views inevitably leads us to the association of life phenomena with a *decrease* in entropy. This in turn, leads to the erroneous (perhaps meaningless) conclusion that life is "anti-entropic" or a "constant struggle" against the Second Law. The fact is that entropy cannot be defined for any living system, and the Second Law, in its entropy formulation does not apply to living systems.

A prize of $ 1,000.00 will be offered to anyone who will be able calculate the entropy of any living creature.

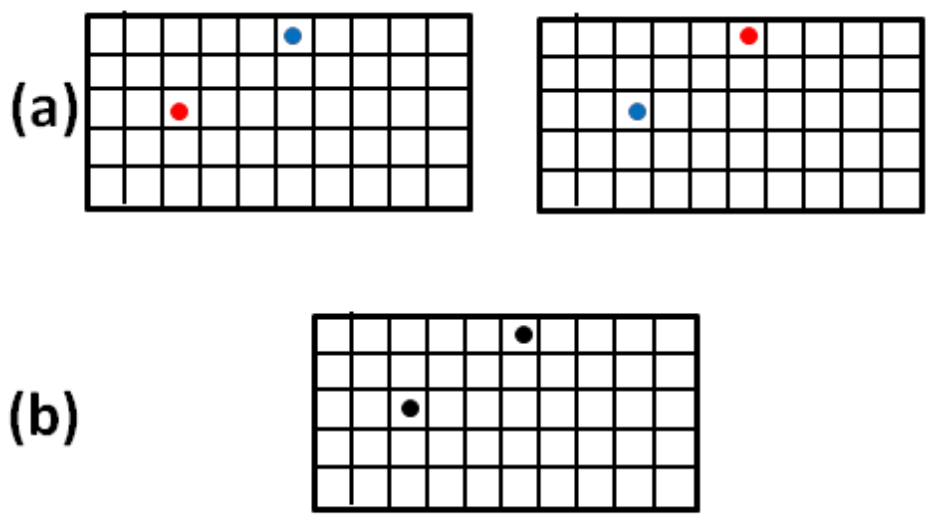

**Figure 1  (a) Two different configurations become
(b)  identical when the particles are indistinguishable**

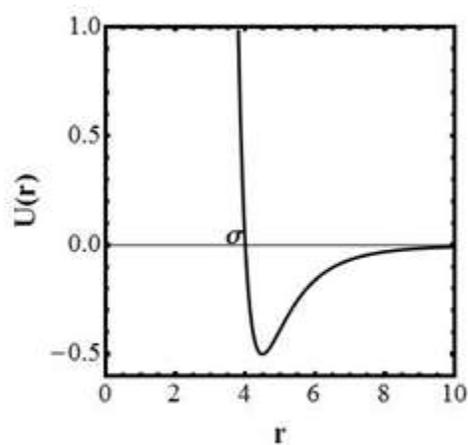

**Figure  2 The general form of
a pair potential between two particles**

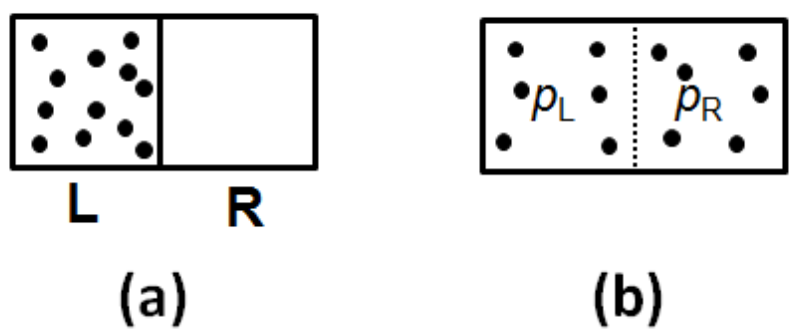

Figure 3 Expansion from *V* to 2 *V*

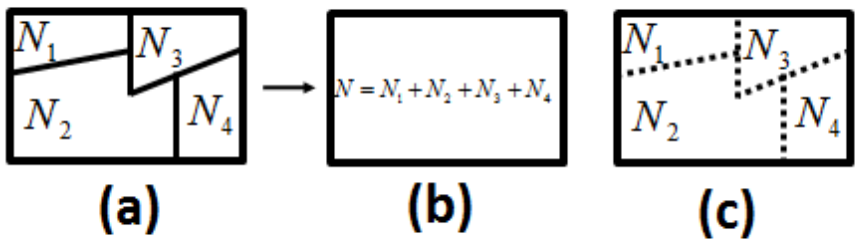

Figure 4. (a) A constrained equilibrium system.
(b) The unconstrained equilibrium state
(c) The same system as in (b) (unconstrained) but with a
distribution of particles as in (a)